\documentstyle[preprint,aps,eqsecnum]{revtex}
\tightenlines
\def\m@thcombine#1#2{%
  \setbox0=\hbox{$#1$}
  \setbox1=\hbox{$#2$}
  \ifdim\wd0>\wd1
    \setbox0=\hbox to\wd1{\hss\box0\hss}
  \else
    \setbox1=\hbox to\wd0{\hss\box1\hss}
  \fi
  \mathop{\vcenter{
    \offinterlineskip\box0\box1}}}
\def\lesim{\m@thcombine<\sim}
\def\gesim{\m@thcombine>\sim}
\begin{document}

\draft

\title{ MASS OF THE $\eta'$ MESON IN THE CHIRAL LIMIT IN THE ZERO MOMENTUM
MODES ENHANCEMENT QUANTUM MODEL OF THE QCD NONPERTURBATIVE VACUUM }

\author{V. Gogohia }

\address{HAS, CRIP, RMKI, Depart. Theor. Phys., P.O.B. 49,
          H-1525, Budapest 114, Hungary \\
      email addresses: gogohia@rmki.kfki.hu and gogohia@rcnp.osaka-u.ac.jp}

\maketitle

\begin{abstract}
 Using the trace anomaly and low energy relations, as well as the
Witten-Veneziano formula for the mass of the $\eta'$ meson, we have developed  
a formalism which makes it possible to express the gluon condensate, the topological susceptibility and  
the mass of the $\eta'$ meson as a functions of the truly nonperturbative vacuum energy density, which is one of the most important characterists of the QCD  
true ground state. It was directly applied to the numerical evaluation of the 
chiral QCD vacuum topological structure within its quantum zero momentum modes 
enhancement model. A rather good agreement with the phenomenological and       
experimental values of the above-mentioned quantities has been achieved. 
With the help of the Witten-Veneziano formula, we derived an absolute lower    
bounds for the pion decay constant and the mass of the $\eta'$ meson in the    
chiral limit. By introducing the most general parametrization of the gluon condensate, we also proposed how the correct $N_f$ (number of flavors) dependence  
of its phenomenologically extracted value could be restored.   
\end{abstract}


\vfill
\eject

\section{Introduction }

It is well known that one of the most important aspects of the famous $U(1)$ problem [1,2] is the large mass of the $\eta'$ meson. It does not vanish in the chiral limit, so the $\eta'$ meson is not the Nambu-Goldstone (NG) boson. In Ref. [3] (see also Ref. [4]) by using the large $N_c$ limit technique the expression for the mass of the $\eta'$ meson was derived, namely

\begin{equation}
m^2_{\eta'} = {2 N_f \over F^2_{\pi}} \chi_t + \Delta ,
\end{equation}
where $\Delta = 2 m^2_{K} - m^2_{\eta}$, $N_f$ is the number of light quarks and $F_{\pi}$ is the pion decay constant. However, the important quantity which enters this formula is the topological density operator (topological susceptibility), $\chi_t$ (for definition see section 3). In the chiral limit it
is screened that is why it is defined for Yang-Mills (YM) fields,
i.e., for pure gluodynamics ($N_f=0$). It is one of the main characteristics
of the QCD nonperturbative vacuum where it measures the fluctuations of the    
topological charge. 

 The precise validity of the Witten-Veneziano (WV) formula (1.1) is, of course,
not completely clear because of its origin. Nevertheless, let us regard it as  
exact for simplicity (in any case we have nothing better than Eq. (1.1). However, there are phenomenological reasons [5,6] as well as some lattice indications
[7] to believe that QCD is close to $SU(\infty)$).       
Using now experimental values of all physical quantities
entering this formula, one obtains that the phenomenological ("experimental")
value of the topological susceptibility is

\begin{equation}
\chi_t^{phen} = 0.001058 \ GeV^4 = (180.36 \ MeV )^4 = 0.1377 \ GeV/fm^3.
\end{equation}
In the chiral limit $\Delta = 0$ since $K^{\pm}$ and $\eta$ particles are
NG excitations.
It is worth noting further that neither the mass of the $\eta'$ meson nor the pion decay constant in the chiral limit cannot exeed their experimental values.
So the WV formula (1.1) provides an absolute lower bounds for the pion decay constant  and the mass of the $\eta'$ meson in the chiral limit, namely

\begin{equation}
854 \leq m^0_{\eta'} \leq 957.77 \ (MeV),
\end{equation}
and

\begin{equation}
83.2 \leq F^0_{\pi} \leq 93.3 \ (MeV),
\end{equation}
respectively. They
should be compared with their experimental values (upper bounds in the previous
expressions). Let us note that the chiral perturbation theory value of the pion
decay constant in the chiral limit, $ F^0_{\pi} = (88.3 \pm 1.1) \ MeV$ [8],  
obviously satisfies these bounds, Eq. (1.4). Recent lattice result [9] (see    
also brief review [7]) for    
the mass of the $\eta'$ meson in the continuum chiral limit is $m^0_{\eta'} =
863(86) \ MeV$, satisfying thus bounds (1.3).

One can conclude in that the mass of the $\eta'$ meson remains large even in   
the chiral limit, which is real problem indeed.
Thus the large mass of the $\eta'$ meson in the chiral limit is due to the phenomenological value of the topological susceptibility.
In other words, it is clear that through the topological susceptibility (i.e.,
via the WV formula (1.1)) the large mass of the $\eta'$ meson even in the
chiral limit is determined by the topological properties of the QCD ground     
state, its nonperturbative vacuum.                                             
It has a very rich dynamical and topological structure
[10-12]. It is a very complicated medium and its dynamical and topological complexity means that its structure can be organized at various levels (quantum, classical). It can contain many different components and ingredients which may
contribute to the truly nonperturbative vacuum energy density (VED). It is well
known that the VED in general is badly divergent [13], however the truly nonperturbative VED is finite, automatically negative and it has no imaginary part   
(stable vacuum). For gauge-invariant definition and concrete examples see      
recent papers [14,15] (for brief description see also sections 2 and 3).       
Precisely this quantity is one of the main characteristics 
of the QCD true ground state and precisely it is related to the                
nonperturbative     
gluon condensate via the trace anomaly relation [16] (section 3) as well   
as to the above-mentioned topological susceptibility via the low energy        
"theorem" (relation) derived by Novikov, Schifman, Vanshtein and Zakharov      
(NSVZ) a long time ego [17] and rederived
quite recently by Halperin and Zhitnitsky (HZ) [18] (section 4).
Let us remind that the truly nonperturbative VED is nothing else but the bag
constant apart from the sign, by definition [13,14,19]. It is much more general
quantity than the string tension because it is relevant for light quarks as    
well.

\section{ZMME quantum model}

 Many models of the QCD vacuum involve some extra classical color field
configurations (such as randomly oriented domains of constant
color magnetic fields, background gauge fields, averaged over
spin and color, stochastic colored background fields, etc.) and
ingredients such as color-magnetic and Abelian-projected
monopoles (see Refs. [10,11,20] and references therein).  The
relevance of center vortices for QCD vacuum by both lattice [21]
and analytical methods [22] was recently investigated as well.
However, the most elaborated classical models are the random and
interacting instanton liquid models (RILM and IILM) of the QCD
vacuum. They are based on the existence of the topologically
nontrivial, instanton-type fluctuations of gluon fields there,
which are nonperturbative, weak coupling limit solutions to the
classical equations of motion in Euclidean space [23] (and
references therein). That is instantons may be $qualitatively$
responsible for the $\eta'$ mass for the first time has been
pointed out by 't Hooft [24]. $Quantitatively$ this problem due to instantons  
was investigated in our previous work [25].                                    

The formalism developed there will be generalized here in order to
investigate this problem (the large mass of the $\eta'$ meson even in the      
chiral limit) $quantititatively$ in quantum field theory as well    
in particular QCD within the above-mentioned zero momentum modes enhancement   
(ZMME) quantum model of the QCD nonperturbative vacuum [26,27].                
        
However, a few general remarks in advance are in order.            
 The quantum part of the VED is, in general, determined  
by the effective potential approach for composite operators [28] (see also     
Ref. [29]). It allows us to  
investigate the nonperturbative QCD vacuum, since in the absence of external  
sources the effective potential is nothing else but the VED. It gives the VED  
in the form of the loop expansion where the    
number of the vacuum loops (consisting of the confining quarks with dynamically
generated quark masses and nonperturbative gluons properly regularized with   
the help of ghosts (if any)) is equal to the power of the Plank                
constant, $\hbar$. As was underlined above, this quantity in general is badly 
divergent at least as fours power of the ultraviolet cutoff (for detail discussion see Shifman's contribution in Ref. [11] as well as our paper [30]).   
 This reflects the fact (situation) that real QCD vacuum  
being nonperturbative, nevertheless contains excitations and fluctuations of  
the gluon field confugurations there which are of pure perturbative character  
and magnitude. In order to deal with its true nonperturbative structure, all   
kind of perturbative contributions should be subtracted from the VED.          
In this way one obtains the truly nonperturbative VED which is precisely the   
one of main characteristics of the QCD true ground state in continuum theory.
This is absolutely similar to lattice approach where by using different 
"smoothing" techniques such as "cooling" [31], "cycling" [32], etc. it is possible to "wash out" all types of the perturbative fluctuations and excitations of
the gluon field configurations from the QCD vacuum in order to deal only with  
its true nonperturbative structure, i.e., free from all kinds of perturbative  
"contamination" ("noise").                             
   
We have already formulated [14,15] a general method how to correctly (in a     
manifestly gauge-invariant way) calculate the truly nonperturbative VED in the 
QCD quantum models
of its ground state using the above-mentioned effective potential approach
for composite operators [28]. The truly nonperturbative VED was defined as
integrated out of the truly nonperturbative part of the full gluon propagator  
over the deep infrared (IR) region (soft momentum region). The nontrivial minimization procedure which can be done only by the two different ways (leading however to the same numerical value (if any) of the truly nonperturbative VED) 
makes it possible to determine the soft cutoff in terms of the corresponding   
nonperturbative scale parameter which is inevitably present in any
nonperturbative model for the full gluon propagator.\footnote{Any deviation of 
the full gluon propagator from the free perturbative one automatically assumes 
its dependence on a characteristic mass scale parameter [30].} The analysis of 
the truly 
nonperturbative Yang-Mills (YM) VED after the scale factorization provides     
an exact criterion for the separation of "stable vs. unstable" vacuum models.  
If the chosen Ansatz for  
the full gluon propagator is a realistic one, then our method uniquely determines the truly nonperturbative YM VED, which is always finite, automatically negative and it has no imaginary part (stable vacuum).     
In the same way (apart from some details concerning the chiral limit physics)  
can be determined the contribution to the truly nonperturbative
VED from confining quark degrees of freedom [26,27]).

The above-briefly-described general method can serve as a test of QCD vacuum different not only quantum, lattice [14,33] but classical models [15,34] as well.
By applying our method to the classical theory, we thereby investigating 
the stability vs. instability of the vacuum of the classical models against   
quantum corrections. However, it is worth emphasizing that the only way to calculate the truly nonperturbative VED in quantum field theory in particular QCD  
from first principles [14,15] is the effective potential approach for composite
operators [28] by substituting there a well-justified ansatz for the full gluon
propagator since an exact solution(s) to its Schwinger-Dyson (SD) equation is  
not yet known. Moreover, it seems to us that it cannot be found in principle   
because of too complicated mathematical structure of the corresponding SD      
equation [2]. Fortunately, however, in order to calculate the truly nonperturbative VED it suffices to know its deep IR asymptotics which for realistic models
usually coincides with its truly nonperturbative part [14]. 

 In our previous works [14,26,27] we have formulated
 a new, quantum model of the QCD ground state (its
nonperturbative vacuum): the ZMME model or simply ZME since we always work     
in the momentum space. It is based on the existence and importance of such     
kind of the nonperturbative, topologically nontrivial quantum excitations of   
the gluon field configurations (due to self-interaction of massless gluons     
only, i.e., without involving
any extra degrees of freedom) which can be effectively, correctly described by 
the $q^{-4}$-type behaviour of the full gluon propagator in the deep IR  
domain. Such strong IR singular behavior of the full gluon propagator can be reffered to as the strong coupling regime, the so-called "infrared slavery" [2]. 
In realistic models for the full gluon propagator its truly nonperturbative part usually coincides with its deep IR asyptotics, emphasizing thus the  
strong intrinsic influence of the IR properties of the theory on its nonperturbative dynamics. By applying the above-described general method to this model,  
the nonperturbative chiral QCD vacuum was found stable, i.e., having a "stationary" state, which can be a manifestation of a possible  existence of a ground  
state in this model. Consequently, the corresponding truly nonperturbative YM  
VED is finite and negative.

Within the ZME quantum model of the QCD ground state,
the truly nonperturbative VED depends on a scale at which the
nonperturbative effects become
important. If QCD itself is confining theory, such a characteristic
scale should certainly exist. The confining quark part of the VED
depends in addition on the constant of integration of the corresponding
quark SD equation in the chiral limit.
 The numerical value of the nonperturbative scale as
well as the above-mentioned constant of integration is obtained
from the bounds (1.4)
for the pion decay constant in the chiral limit by implementing a
physically well-motivated scale-setting scheme [26,27].\footnote{In nonperturbative QCD all numerical results depend on the numerical value of the nonperturbative (characteristic) scale and not on $\alpha_s$ as in phenomenological and   
perturbative QCD. This 
is due to phenomenon of "dimensional transmutation" in field theories with spontaneous symmetry breaking [35,36] (see also discussion in Ref. [30]).} We have
obtained the following numerical results for the truly nonperturbative VED    
due to ZME quantum model:

\begin{eqnarray}
\epsilon_{ZME} &=& - (0.01413 - N_f 0.00196)  \ GeV^4, \nonumber\\
\epsilon_{ZME} &=& - (0.009 - N_f 0.00124)  \ GeV^4,
\end{eqnarray}
where, obviously, the first and second values are due to upper and lower bounds
in Eq. (1.4), respectively.                                                  
Here and further on below $N_f$ is the number of light flavors. We see that    
the confining quark part is approximately one order of magnitude less than     
the YM part and it is of opposite sign. 

It is instructive to compare our values (2.1) with those which are due to      
instantons in the chiral limit [25]. Within the above-mentioned RILM, for 
dilute ensemble, the truly nonperturbative VED is $\epsilon_I = - (b/4) \times n_0$. Here $b= 11 - (2/3)N_f$ is the first coefficient of the $\beta$ function
(see below) and $n_0$ is the instanton number density in the chiral limit [23,25]. Due to all reasonable estimates of this quantity (which follows from phenomenology or lattice approach) its numerical value cannot exceed its phenomenological value, i.e., $n_0 \leq n = 1.0 \ fm^{-4}$. Thus, at maximum the truly     
nonperturbative VED due to RILM is 

\begin{equation}
\epsilon_I = - (0.004179 - N_f 0.00025)  \ GeV^4.
\end{equation}
It is clear that our values (2.1) approximately one order of magnitude bigger  
than those of instantons can provide at all (2.2).

One of the main purposes in this paper is to generalize a     
formalism, developed earlier in our paper [25], in order to     
directly calculate the gluon condensate as a function of the truly             
nonperturbative VED, the topological
susceptibility and the mass of the $\eta'$ meson in the chiral
limit as a functions of the truly nonperturbative YM VED. It is based on using
the trace anomaly (section 3) and low energy (section 4) relations, as well as 
the WV formula for the mass of the $\eta'$ meson in the chiral limit (section  
5). It was directly applied to the ZMME quantum model of the QCD nonpertubative
vacuum. In section 6 we present our discussion and conclusions. Our numerical  
results are shown in Tables I-III.

\section{ The trace anomaly relation }

The truly nonperturbative VED is important in its own right as one of the
main characteristics of the QCD nonperturbative vacuum.
Furthermore it assists in estimating such an important
phenomenological parameter as the gluon condensate, introduced in
the QCD sum rules approach to resonance physics [37]. The famous
trace anomaly relation [16] in the general case (nonzero current
quark masses $m_f^0$) is

\begin{equation}
\Theta_{\mu\mu} = {\beta(\alpha_s) \over 4 \alpha_s}
G^a_{\mu\nu} G^a_{\mu\nu}
+ \sum_f m_f^0 \overline q_f q_f.
\end{equation}
where $\Theta_{\mu\mu}$ is the trace of the energy-momentum tensor
and $G^a_{\mu\nu}$ being the gluon field strength tensor while
$\alpha_s = g^2/4 \pi$.
Sandwiching Eq. (3.1) between vacuum states and on account of the
obvious relation $\langle{0} | \Theta_{\mu\mu} | {0}\rangle = 4
\epsilon_t$, one obtains

\begin{equation}
4 \epsilon_t = 
\langle{0} | {\beta (\alpha_s)  \over 4 \alpha_s} G^a_{\mu\nu} G^a_{\mu\nu} |
 {0}\rangle
+ \sum_f m^0_f \langle{0} | \overline q_f q_f | {0}\rangle,
\end{equation}
where $\langle{0}| \overline q_f q_f | {0} \rangle$ is the quark condensate    
and $\epsilon_t$ is the sum of all possible independent, truly               
nonperturbative contributions to the VED (the total VED). In quantum theory it 
consists of two parts:
the truly nonperturbative gluon part properly regularized with the help of 
ghosts (if any), $\epsilon_g$, the so-called YM part at $N_f=0$, and confining 
quark part with dynamically generated quark masses, $\epsilon_q$, i.e., 

\begin{equation}
\epsilon_t = \epsilon_g + N_f \epsilon_q.
\end{equation} 

In general case it is impossible to use the weak coupling limit solution to 
the $\beta$ finction, so it is convenient to introduce the gluon condensate as 
follows:

\begin{equation}
\bar G_2 \equiv \langle \bar G^2 \rangle \equiv      
- \langle {\beta(\alpha_s) \over 4 \alpha_s } G^2 \rangle \equiv
- \langle{0} | {\beta(\alpha_s) \over 4 \alpha_s } G^a_{\mu\nu} G^a_{\mu\nu}   
| {0}\rangle.
\end{equation}
This is nothing else but the most general parametrization of the gluon correlation function, $\langle {0} | G^a_{\mu\nu} G^a_{\mu\nu} | {0}\rangle$ (for details, see sect. VI below) which comes from the trace anomaly ralation (3.2)      
itself.  
If confinement happens then the $\beta$ function is always in the domain of    
attration (i.e., always negative) without IR stable fixed point [2]. Thus the  
gluon condensate, parametrized as in Eq. (3.4), is always positive as it should
be. The trace anomaly relation (3.2) then can be written as follows:

\begin{equation}
4 \epsilon_t = - \bar G_2 + 
 \sum_f m^0_f \langle{0} | \overline q_f q_f | {0}\rangle.
\end{equation}
Since the gluon condensate $\bar G_2$ is always positive and finite    
while quark condensate is always negative and also finite, then the truly nonperturbative total VED $\epsilon_t$ is also always negative and finite without   
having imaginary part since $\bar G_2$ and quark condensate are real finite    
numbers as it has been emphasized above. As was underlined in the preceding    
section, the confining quark part of the total vacumm energy density           
$\epsilon_q$ (which is always positive) as usual an order of magnitude less than the corresponding YM part $\epsilon_g$ (which is always negative), so the total truly nonperturbative VED $\epsilon_t$ is always negative in complete agreement with Eq. (3.5). On account of Eq. (3.3), it gives the gluon condensate     
(3.4) as a function of $N_f$ if somebody knows, of course, how to calculate    
$\epsilon_t$ and quark condensate from first principles (see below and our 
papers [14,26,27]).

 In the chiral limit ($m^0_f=0$) things drastically simplify. 
From Eq. (3.5) one obtains, $\bar G_2^0 = - 4 \epsilon_t^0$,
indeed, where superscript "0" means the chiral limit. Thus in this limit       
the truly     
nonperturbative total VED is nothing else but the gluon condensate apart from  
the overall numerical factor.        
In what follows we will saturate the total VED in this equation by our values  
(2.1), i.e., to put  $\epsilon_t^0 = \epsilon_{ZME}  + ...$. The numerical     
results are shown in Table I.

\section{ The topological susceptibility}

One of the main characteristics of the QCD nonperturbative vacuum is
 the topological density operator (topological susceptibility)
in gluodynamics ($N_f=0$) [3]

\begin{equation}
\chi_t = \lim_{q \rightarrow 0} i \int d^4x\, e^{iqx} {1 \over N_c^2} \langle{0} | T \Bigl\{ {q(x) q(0)} \Bigr\} | {0} \rangle,
\end{equation}
where $q(x)$ is the topological charge density, defined as
$q(x) = (\alpha_s / 4 \pi) F (x) \tilde{F} (x) = (\alpha_s / 4 \pi)
F^a_{\mu \nu} (x) \tilde {F}^a_{\mu \nu}(x)$
and $\tilde {F}^a_{\mu \nu}(x) = (1 / 2)
\epsilon^{\mu \nu \rho \sigma} F^a_{\rho \sigma} (x)$
is the dual gluon field strength tensor while $N_c$ is the number of different 
colors. In the definition of the topological
susceptibility (4.1) it is assumed that the corresponding regularization
and subtraction of all types of the perturbative contributions have been       
already done in order Eq. (4.1) to stand for the renormalized, finite and      
the truly nonperturbative topological susceptibility (see Refs. [3,17,18,38]).
The anomaly equation in the WV notations is

\begin{equation}
\partial_{\mu} J_5^{\mu} = N_f (2 / N_c) (\alpha_s / 4 \pi) F \tilde{F}.
\end{equation}

The topological susceptibility
can be related to the nonperturbative gluon condensate via the
low energy "theorem" in gluodynamics proposed by NSVZ [17] (by using the       
dominance of self-dual fields hypothesis in the YM vacuum) as follows:

\begin{equation}
\lim_{q \rightarrow 0} i \int d^4x\, e^{iqx} \langle{0} | T \Bigl\{ {\alpha_s \over 8 \pi} G \tilde{G} (x)
{\alpha_s \over 8 \pi} G \tilde {G}(0) \Bigr\} | {0}
\rangle = - \xi^2 \langle {\beta(\alpha_s) \over 4 \alpha_s } G^2 \rangle.
\end{equation}  
Quite recently it was discussed by HZ in Ref. [18] (see also references        
therein) who noticed that it is not precisely a Ward identity, but rather 
is a relation between the corresponding correlation functions, indeed. That is 
why in what follows we call Eq. (4.3) as low energy relation or NSVZ-HZ        
relation. Thus there exist two proposals how to fix the numerical value of the 
coefficient $\xi$. The value $\xi = 2/ b$, where here and everywhere below  
(apart from section 6)   
$b=11$, was suggested by NSVZ [17] who used the above-mentioned dominance of 
self-dual fields hypothesis in the YM vacuum. A second one , $\xi = 4/ 3b$, was
advocated very recently by HZ using a one-loop connection between the conformal
and axial anomalies in the theory with auxiliarly heavy fermions [18] (and     
references therein). For completeness we will use both values for the $\xi$    
parameter.\footnote{In the weak coupling limit solution to the $\beta$ function
(see, expression (6.2) below), the NSVZ-HZ low energy relation (4.3) for the   
NSVZ value of the $\xi$ parameter is reduced (as it should be) to that which   
was used in our previous paper [25] where the instanton vacuum in the chiral   
limit in various modifications was investigated. At the same time, the functional dependence of the topological susceptibility on the vacuum energy density   
is not, of course, changed (compare expression (3.5) of Ref. [25] with Eq. (4.5) below, on account of the corresponding values of the $\xi$ parameter).}    
Let us note only that there exists an obvious relation between these two       
values, namely $\xi_{HZ} = (2/3) \xi_{NSVZ}$.                                  
                 
The anomaly equation in the NSVZ-HZ notations is

\begin{equation}
\partial_{\mu} J_5^{\mu} = N_f (\alpha_s / 4 \pi) G \tilde{G},
\end{equation}
with $N_f=3$. Thus in order to get the topological susceptibility in the WV form from the relation (4.3), it is necessary to make a replacement in its left hand side as follows: $G \tilde{G} \longrightarrow (2 / N_c)  F \tilde{F}$ in accordance with the anomaly equations (4.2) and (4.4). Then the WV topological susceptibility (4.1) finally becomes    

\begin{equation}
\chi_t  = - \xi^2 \langle {\beta(\alpha_s) \over 4 \alpha_s } G^2 \rangle =
- (2 \xi)^2 \epsilon_{YM},
\end{equation}
where the second equality comes from Eqs. (3.4) and (3.6) by denoting the      
truly nonperturbative VED at $N_f=0$ as $\epsilon_{YM}$. 
The significance of this formula is that it gives the topological
susceptibility as a function of the truly nonperturbative VED for pure gluodynamics, $\epsilon_{YM}$. For numerical results in the NSVZ mode see Table II.    
In order to obtain numerical results for the topological susceptibility in the 
HZ mode, it suffices in this     
Table to multiply numbers in $GeV$ units by factor $(2/3)^2$ and numbers in
$MeV$ units by factor $\sqrt{2/3}$.

\section{ The $U(1)$ problem}

 The topological susceptibility (4.1) assists
in the resolution of the above-mentioned $U(1)$
problem [1,2]
via the WV formula for the mass of the $\eta'$ meson (1.1).
 Within our notations it is expressed (in the chiral limit) as follows:
$f^2_{\eta'} m^2_{\eta'} = 4 N_f \chi_t$,
where $f_{\eta'}$ is the $\eta'$ residue defined in general as
$\langle {0}| \sum_{q=u,d,s} \overline q \gamma_{\mu} \gamma_5 q | {\eta'}
\rangle = i \sqrt{N_f} f_{\eta'} p_{\mu}$ and $\langle {0}| N_f {\alpha_s \over 4 \pi} F \tilde{F} | {\eta'} \rangle = (N_c \sqrt{N_f} / 2) f_{\eta'} m^2_{\eta'}$ [3].
 Using the normalization relation $f_{\eta'} = \sqrt{2} F^0_{\pi}$, one
 finally obtains

\begin{equation}
F^2_{\pi}m^2_{\eta'} = 2 N_f \chi_t.
\end{equation}
Eq. (4.5) then implies

\begin{equation}
m^2_{\eta'} = - 2 N_f \Bigl( {2 \xi \over F_{\pi} } \Bigr)^2 \epsilon_{YM},
\end{equation}
which expresses the mass of the $\eta'$ meson as a function of the truly nonperturbative YM VED.
In previous expressions we omit for simplicity the
superscript "0" in the pion decay constant as well as in $m^2_{\eta'}$. 
Our numerical results for the mass of the $\eta'$ meson in the chiral limit    
(5.2) are shown in Table III.

\section{Discussion and conclusions }

\subsection{$N_f$ dependence of the gluon condensate }

Let us emphasize, that the general parametrization of the gluon condensate,   
introduced in Eq. (3.4), can be formally applied to $any$ $\beta$ function: to 
its weak coupling limit solution which we certainly know (see below) or to its 
strong coupling limit one which we certainly do not know yet. Also it is       
remains relevant whether one considers the chiral limit case or not. 
In phenomenology, however, the oftenly used parametrization of the gluon      
condensate is [37]                                                             
              
\begin{equation}
 G_2  \equiv \langle G^2 \rangle \equiv     
\langle {\alpha_s \over \pi } G^2 \rangle \equiv
\langle{0} | {\alpha_s \over \pi } G^a_{\mu\nu} G^a_{\mu\nu} |{0}\rangle, 
\end{equation}
which in what follows we will call standard (or phenomelogical) parametrization
of the gluon condensate. Of course, the gluon condensate in principle can be parametrized by different ways. Here it makes sense to remind that by parametrization it is understood which numerical factor ($\alpha_s / \pi, \ \alpha_s$, etc.) is chosen to be associated with the gluon correlation function 
$\langle{0} | G^a_{\mu\nu} G^a_{\mu\nu} |{0}\rangle$.
By comparing these two parametrizations of the same gluon correlation function,
it becomes clear that the  general parametrization $\bar G_2$, given in        
Eq. (3.4), is useful in 
the nonperturbative calculations from first principles as it has been explained
above. At the same time, it is useless in phenomenology since we do not know   
the $\beta$ function there. However, in phenomenology in many important cases  
it is legitimated to use the weak coupling limit    
solution to the $\beta$ function,

\begin{equation}
\beta(\alpha_s) = - b {\alpha_s^2 \over 2 \pi} + 0(\alpha_s^3), \quad b = 11 - 
{2 \over 3} N_f,
\end{equation}
for example in instanton calculus [23,25]. Then the general parametrization    
(3.4) is reduced to the phenomenological one, Eq. (6.1), as follows: 

\begin{equation}
\bar G_2  = {b \over 8} \times  G_2 = (1.375 - 0.083 N_f) \times G_2.
\end{equation} 
This relation makes it possible to resolve the old-standing problem in QCD phenomenology. It is well known that the phenomenologically extracted value of     
the gluon condensate does not explicitly depend on $N_f$. It is usually       
nonexplicitly assumed that it is relevant for all $N_f$ by maching from heavy  
to light quarks [37]. Though in general it should be $N_f$ explicitly dependent
quantity as it follows from the trace anomaly relation (3.5) on account of     
Eq. (3.3). This is true in the weak coupling regime and in the chiral limit    
as well.

Our proposal is to consider (to interpret) $\bar G_2$  as 
a real phenomenological condensate which on account of any numerical value of  
the standard gluon condensate $G_2$, through the relation (6.3), becomes explicitly dependent on $N_f$, indeed. In other words, our proposal allows one to    
restore the explicit $N_f$ dependence of the phenomenologically determined     
gluon condensate. The relation (6.3) in fact prescribes which numerical  
factor, explicitly depending on $N_f$, should be multiplied on $G_2$ in  
order to fix the correct $N_f$ dependence of the gluon condensate in           
phenomenology. This interpretation is in agreement with the trace anomaly    
relation, of course.  

Another problem is the numerical value of the standard gluon condensate $G_2$  
itself. It can be taken either from phenomenology or lattice simulations 
(see below). The phenomenological analysis of QCD sum rules for the       
standard gluon condensate implies, $G_2 \simeq  0.012 \ GeV^4$,
which can be changed within a factor of two [37]. However, it has been         
already pointed out [39] that QCD sum rules substantially
underestimate the value of the standard gluon condensate.
The most recent phenomenological calculation of the standard gluon condensate  
is given by Narison in Ref. [40], where a brief review of many previous        
calculations is also presented.
His analysis leads to the update average value as
$G_2  = (0.0226 \pm 0.0029) \ GeV^4$.
In Ref. [41] from the families of $J/ \Psi$ and $\Upsilon$ mesons              
substantially larger values were recently derived, namely
$0.04 \leq G_2 \leq 0.105 \ (GeV^4)$.
Comparable with these bounds, the values for the standard gluon condensate
were reported in recent lattice simulations [42].
  
 Obviously that the standard gluon condensate in the chiral limit cannot be equal to its any phenomenological (empirical) value. Apparently it should be      
less, i.e., $G^0_2 \simeq \nu \times G_2$, where $\nu < 1$ is some real number.
Then the relation (6.3) in the chiral limit becomes
$\bar G^0_2 = (b /8) \times  G_2^0 \simeq (\nu b / 8) \times   
G_2 = (1.375 - 0.083 N_f) \times \nu \times G_2$.
 In Ref. [17] it has been argued indeed that the gluon condensate in the chiral
limit is approximately two times less than its any phenomenological (empirical)
value, $G^0_2 \simeq 0.5 G_2$, i.e., $\nu \simeq 0.5$. Then the previous relation becomes 
$\bar G^0_2 \simeq (b / 16) \times  G_2 = (0.6875 - 0.0415 N_f) \times G_2$,
whatever numerical value of $G_2$ is.

Our values for the gluon condensate (3.4) in the chiral limit (Table I)
substantially larger than it is possible to estimate from phenomenology at all 
with the help of the above-derived relations. This difference is not only due
to different physical observables as was noticed in Ref. [23].  It seems to us 
that it reflects rather different underlying physics as well. Our gluon condensate (which was calculated from first principles) is the strong coupling regime 
result and reflects the nontrivial topology  
of the true QCD vacuum where quantum excitations of the gluon fields play      
an important role. Precisely these types of gluon field configurations are mainly responsible for quark confinement and dynamical breakdown of chiral symmetry
[14,26,27]. At the same time, phenomenologically extracted gluon condensate    
being
the weak coupling limit result can be associated with classical instanton-type fluctuations in the QCD vacuum which by themselves do not confine quarks       
[32,43,44].     

Concluding this subsection, let us note that the confining quark condensate    
contribution to
the trace anomaly relation (3.5) vanishes in the chiral limit.  However, due to
all reasonable estimates of light quark masses, numerically its contribution
is at $20\%$ and thus comparable with the systematic error in any phenomenological determination of the standard gluon condensate itself [37,45].

\subsection{Some general remarks on the trace anomaly relation }

It is widely believed that the truly nonperturbative VED is determined by the  
trace anomaly relation (3.2) or equivalently (3.5). This may be so in phenomenology but not, of course, in nonperturbative QCD, where the situation is completely opposite (see discussion below).  If by some phenomenological analysis it  
is possible to extract numerical values of the gluon and quark condensates then
substituting them into the right hand side of the trace anomaly relation (3.5),
one is able to estimate the total VED, indeed. However, this number such       
obtained tells a little (or even nothing) about detail      
structure of the QCD vacuum. It is not surprising since the gluon and quark condensates being its average (global) characteristics cannot account for its 
microscopic, detail structure.                                       

The quantity which is responsible for 
its detail structure is precisely the total truly nonperturbative VED which,
by definition, is the sum of all possible, independent contributions. In this  
way it refrects the complexity and veriety of the quantum, dynamical degrees of
freedom (quarks and gluons) in the QCD vacuum. It may cotain contributions even
from classical field configurations (instantons, for example). Precisely the   
detail structure of the QCD vacuum determines its global characteristics, and  
not vice versa.   
We have already explained in section 2 how the truly nonperturbative VED should
be calculated from first principles in nonperturbative QCD. It is well known   
how to calculate from first principles the quark condensate as integrated out  
of the trace of the truly nonperturbative quark propagator over the deep IR    
region. Then the trace anomaly relation (3.5) becomes simply an algebraic equation for calculating the gluon condensate and in the chiral limit it is nothing 
else but the truly nonperturbative VED itself.          

The contribution to the right hand side of 
the trace anomaly relation (3.5) from the confining quark condensate vanishes  
in the chiral limit. However, it does not mean that the contribution from 
confining quarks with dynamically generated quark masses also vanishes in the  
left hand side of the trace anomaly relation. Even in the chiral limit the     
vacuum quark loops provide nonzero contribution into the total VED. Thus this  
quantity contains much more information about QCD vacuum than its global       
chracteristics can provide at all. Reflecting its detail structure the total   
VED is the main characteristics of the QCD nonperturbative vacuum and thereby  
is responsible for its global characteristics as well.

\subsection{Summary}

Using the trace anomaly relation (3.2), The NSVZ-HZ
low energy relation (4.3) and WV formula for the mass of the $\eta'$ meson     
(5.2) in the chiral limit, we have developed a well-justified formalism in the 
most general form in order to express the topological susceptibility, the gluon
condensate and the mass of the $\eta'$ meson in the chiral limt as a functions 
of the truly nonperturbative VED. The crucial role in this belongs to the      
NSVZ-HZ low-energy relation (4.3). It allows one to relate the important   
topological quantities such as the gluon condensate, the truly nonperturbative 
VED, the topological susceptibility, etc. to each other in a self-consistent   
way by providing well-justified coefficients between them.                     

This formalism has been   
immediately applied to the investigation of the chiral QCD vacuum structure    
within its quantum ZMME model. Our numbers are collected  
in Tables I-III. Our values for the topological susceptibility in the NSVZ mode
(Table II) slightly overestimate its phenomelogical value while in the HZ mode
underestimate it since recalling $\xi_{HZ} = (2/3) \xi_{NSVZ}$ (see also text  
at the very end of section 4). The same situation takes place with the mass of 
the $\eta'$ meson in the chiral limit         
(Table III). However, it is worth emphasizing that the precise validity neither
of the WV formula (1.1) nor the NSVZ-HZ  
low energy relation (4.3) is not known. That is why we cannot address the question of systematic error bars one might assign to the final numerical values summarized in Tables II and III.                                       
The topological susceptibility to leading order in the large $N_c$ limit is of 
order $N_c^0$. The next-to-leading correction of order $N_c^{-1}$, however     
maight be not very small at $N_c=3$ which was   
used in our calculations.                                                      

Any way, the ZMME 
values of the truly nonperturbative VED (2.1) are of the necessary order of magnitude to account for the phenomenological value of the topological susceptibility and therefore to saturate the large mass of the $\eta'$ meson.          
In other words, the IR singularities likely to be presented in the QCD nonperturbative vacuum and summarized by the corresponding behavior of the full gluon  
propagator in the deep IR domain are mainly responsible for the large mass of 
the $\eta'$ meson and consequently for the phenomenological value of the topological susceptibility. Precisely this type of gluon field configurations is closely related to quark confinement as well [14,26,27]. 
Also our proposal how to restore the correct $N_f$ dependence of the phenomenologically extracted value of the standard gluon condensate seems to be         
important in general and for instanton calculus [23] in particular. It is well 
known that the gluon condensate due to instantons does not explicitly depend   
on $N_f$ [23,25,37].  

 The author is grateful to H. Toki for many interesting 
and useful discussions on these topics during his stay at RCNP. It is also a   
pleasure to thank Gy. Kluge for many useful remarks and help.

\vfill

\eject

\vfill

\eject

\begin{table}
\caption{ZMME model values for the gluon condensate. }
\begin{center}
\renewcommand{\arraystretch}{1.5}
\begin{tabular}{lrrrrr}
$F^0_{\pi} \ (MeV)$ &$\bar G_2^0 \ (GeV^4)$ & $N_f=0$ & $N_f=1$ & $N_f=2$ & $N_f=3$ \\
\hline
93.3 &  & $0.05652$ & $0.04936$ & $0.04084$ & $0.0330$ \\
83.2 &  & $0.0360$ & $0.03104$ & $0.02608$ & $0.02112$  \\
\end{tabular}
\renewcommand{\arraystretch}{1.0}
\end{center}
\end{table}

\begin{table}
\caption{ZMME model values for the topological susceptibility.}
\begin{center}
\renewcommand{\arraystretch}{1.5}
\begin{tabular}{lrr}
$F^0_{\pi} \ (MeV)$ & $\chi_t \ (GeV^4)$ & $\chi_t^{1/4} \ (MeV)$ \\
\hline
93.3 & 0.00187 & 207.9\\
83.2 & 0.00119 & 185.7 \\
\end{tabular}
\renewcommand{\arraystretch}{1.0}
\end{center}
\end{table}

\begin{table}
\caption{ZMME model values for the mass of the $\eta'$ meson     
in the chiral limit ($MeV$ units).}
\begin{center}
\renewcommand{\arraystretch}{1.5}
\begin{tabular}{lrr}
$F^0_{\pi}$ & $m_{\eta'}^0 \ (NSVZ)$ & $m_{\eta'}^0 \ (HZ)$ \\
\hline
93.3 & 1135 & 756.86 \\
83.2 & 1015  & 677 \\
\end{tabular}
\renewcommand{\arraystretch}{1.0}
\end{center}
\end{table}


\begin{references}
\bibitem{1}
   S. Weinberg, Phys. Rev. D 11 (1975) 3583
\bibitem{2}
   W. Marciano, H. Pagels, Phys. Rep. C 36 (1978) 1; \\
   G.A. Christos, Phys. Rep. 116 (1984) 251
\bibitem{3}
   E. Witten, Nucl. Phys. B 156 (1979) 269
\bibitem{4}
   G. Veneziano, Nucl. Phys. B 159 (1979) 213
\bibitem{5}
   G. 't Hooft, Nucl. Phys. B 72 (1974) 461
\bibitem{6}
   E. Witten, Nucl. Phys. B 160 (1979) 57
\bibitem{7}
   M. Teper, Nucl. Phys. B (Proc. Suppl.) 83-84 (2000) 146; \\
   A. Hart, M. Teper, hep-lat/0009008
\bibitem{8}
   P. Gerber, H. Leutwyler, Nucl. Phys. B 321 (1989) 387
\bibitem{9}
   A. Ali Khan et al., (CP-PACS), hep-lat/9909045
\bibitem{10}
   P. van Baal (Ed.). Confinement, Duality and Nonperturbative Aspects of QCD.
NATO ASI Series B: Physics, vol. 368
\bibitem{11}
   K-I. Aoki, O. Miymura, T. Suzuki (Eds.). Non-Perturbative QCD. Structure of the QCD Vacuum,
    Prog. Theor. Phys. Suppl. 131 (1998) 1
\bibitem{12}
   V.N. Gribov, Eur. Phys. Jour. C 10 (1999) 91
\bibitem{13}
   E.V. Shuryak, Phys. Rep. 115 (1984) 151
\bibitem{14}
   V. Gogohia, Gy. Kluge, Phys. Rev. D 62 (2000) 076008
\bibitem{15}
   V. Gogohia, Gy. Kluge, Phys. Lett. B 477 (2000) 387 
\bibitem{16}
   R.J. Crewther, Phys. Rev. Lett. 28 (1972) 1421; \\
   M.S. Chanowitz, J. Ellis, Phys. Rev. D 7 (1973) 2490; \\
   J.C. Collins, A. Duncan, S. D. Joglecar, Phys. Rev. D 16 (1977) 438
\bibitem{17}
   V.A. Novikov, M.A. Shifman, A.I. Vainshtein, V.I. Zakharov,
    Nucl. Phys. B 191 (1981) 301
\bibitem{18}
   I. Halperin, A. Zhitnitsky, Nucl. Phys. B 539 (1999) 166
\bibitem{19}
   M.S. Chanowitz, S. Sharpe, Nucl. Phys. B 222 (1983) 211
\bibitem{20}
   J. Hosek, G. Ripka, Z. Phys. A 354 (1996) 177; \\
   Yu.A. Simonov, Phys. Usp. 166 (1996) 337; \\
   H.G. Dosch, Prog. Part. Nucl. Phys. 33 (1994) 121
\bibitem{21}
   T.G. Kovacs, E.T. Tomboulis, Nucl. Phys. B (Proc. Suppl.) 73 (1999) 566; \\ 
   L. Del Debbio, M. Faber, J. Greensite, S. Olenik, Nucl. Phys. B (Proc. Suppl.) 63 (1998) 552; \ A. Ivanov et al., private communication; \\
  P. de Forcrand, M. D'Elia, Phys. Rev. Lett. 82 (1999) 4582
\bibitem{22}
   J.M. Cornwall, Phys. Rev. D 61 (2000) 085012
\bibitem{23}
   T. Schafer, E.V. Shuryak, Rev. Mod. Phys. 70 (1998) 323
\bibitem{24}
   G. 't Hooft, Phys. Rev. Lett. 37 (1976) 8
\bibitem{25}
   V. Gogohia, to appear in Phys. Lett. B, see also hep-ph/0005302
\bibitem{26}
   V. Gogohia, Gy. Kluge,  H. Toki, T. Sakai, Phys. Lett. B 453 (1999) 281
\bibitem{27}
   V. Gogohia, H. Toki, T. Sakai, Gy. Kluge, Inter. Jour. Mod. Phys. A B 15 (2000) 45
\bibitem{28}
   J.M. Cornwall, R. Jackiw, E. Tomboulis, Phys. Rev. D 10 (1974) 2428
\bibitem{29}
   A. Barducci et al., Phys. Rev. D 38 (1988) 238; \\
   R. W. Haymaker, Riv. Nuovo Cim. 14 (1991) 1; \\
   K. Higashijima, Prog. Theor. Phys. Suppl. 104 (1991) 1
\bibitem{30}
   V. Gogohia, Phys. Lett. B 485 (2000) 162
\bibitem{31}
   D.A. Smith, M.J. Teper, Phys. Rev. D 58 (1998) 014505; \\
   M.-C. Chu, J.M. Grandy, S. Huang, J.W. Negele, Phys. Rev. D 49 (1994) 6039
\bibitem{32}
   T. DeGrand, A. Hasenfratz, T.G. Kovacs, Nucl. Phys. B 505 (1997) 417; \\
   A. Hasenfratz, C. Nieter, Nucl. Phys. B (Proc. Suppl.) 73 (1999) 503
\bibitem{33}
   M. Prisznyak, hep-ph/0011029
\bibitem{34}
   V. Gogohia, M. Prisznyak, Phys. Lett. B 494 (2000) 109
\bibitem{35}
   S. Coleman, E. Weinberg, Phys. Rev. D 7 (1973) 1888; \\
   D.J. Gross, A. Neveu, Phys. Rev. D 10 (1974) 3235
\bibitem{36}
   F. Wilczek, Proc. Inter. Conf., QCD - 20 Years Later, Aachen, June 9-13, v. 1, p.16
\bibitem{37}
   M.A. Shifman, A.I. Vainshtein, V.I. Zakharov, Nucl. Phys. B 147 (1979) 385, 448  
\bibitem{38}
   E. Meggiolaro, Phys. Rev. D 58 (1998) 085002; \\
   G. Gabadadze, Phys. Rev. D 58 (1998) 094015
\bibitem{39}
   J.S. Bell, R.A. Bertlmann, Nucl. Phys. B 177 (1981) 218; \\
   R.A. Bertlmann et al., Z. Phys. C 39 (1988) 231
\bibitem{40}
   S. Narison, Phys. Lett. B 387 (1996) 162
\bibitem{41}
   B.V. Geskenbein, V. L. Morgunov, Yad. Fiz. 58 (1995) 1980
\bibitem{42}
   M. Camprostini, A. Di Giacomo, Y. Gunduc, Phys. Lett. B 225 (1989) 393; \\
   B. Alles et al., Phys. Rev. D 48 (1993) 2284
\bibitem{43}
   R.C. Brower, D. Chen, J.W. Negele, E.V. Shuryak, Nucl. Phys. B (Proc. Suppl.) 73 (1999) 512; \ M.I. Polikarpov, A.I. Veselov, Nucl. Phys. B 297 (1988) 34
\bibitem{44}
   C.G. Callan, Jr., R. Dashen, D.J. Gross, F. Wilczek, A. Zee, Phys. Rev. D 18
(1978) 4684
\bibitem{45}
   M. Schaden, Phys. Rev. D 58 (1998) 025016
\end{references}
\end{document}